\documentstyle[eqsecnum, aps, prbbib]{revtex}
\def\7x7{Si(111)$7\!\times\!7$}

\def\mic{${\rm \mu}$}
\begin{document}
\draft
%
\title{Electron microscopy of the strain on the \7x7 surface induced by the 
STM tip
}
\author{
Y. Naitoh, K. Takayanagi, Y. Oshima and H. Hirayama}
\address{
and Technology Corporation (JST), \\
Department of Materials Science and Engineering, Interdisciplinary Graduate 
School of Science and Engineering, Tokyo Institute of Technology\\
4259 Nagatsuda-cho, Midori-ku, Yokohama 226-8502, Japan}
\date{\today}
\maketitle
\begin{abstract}
The \7x7 surface was observed by reflection electron microscopy (REM) and 
scanning tunneling microscopy (STM) simultaneously in an ultra-high vacuum 
electron microscope. The distance between the STM tip and the Si surface 
was detected from the REM image, which showed the real and the mirror image 
of the tip. We approached the tip to the surface or retracted from the 
surface by a piezo drive to observe the strain induced on the \7x7 surface 
as a function of the tip-surface distance. This investigation was done with 
and without the bias voltage between the tip and the substrate. With bias 
voltage of 1.0\,V on the sample, the tip was approached to 1.6\,nm above 
the sample surface for the tunneling current of 0.8\,nA, no detectable 
order of strain ($\sim{}10^{-4}$) was induced on the sample surface. When 
the bias decreased within the range of $-0.3$\,V\,$\sim$\,$+0.5$\,V, the 
surface was compressed over the Si surface area of 100\,nm.
Without the bias voltage, tensile and compressive strain was detected as 
the tip-surface distance changed from attractive to the repulsive 
interaction regime. The strain field extended over 50\,nm\,$\sim$\,140\,nm, 
and the force became neutral at the tip-substrate distance of 0.45\,nm.

\end{abstract}


\narrowtext

\section{Introduction}

Scanning tunneling microscope (STM) and atomic force microscopy (AFM) has 
been used not only as a mean of microscopy, but also as a tool for 
nanofabrication, nanoprobing, or manipulation of atoms and molecules on 
surfaces.\cite{eigler,avouris,aono,bartel,dujardin}  The techniques attract 
much interest in the fields of nano-devices and molecular electronics. In 
these techniques modification by tips \cite{tmmeyer,avouris2} or the 
migration of surface atoms caused by high electric field 
\cite{whitman,nakayama} are utilized. Dynamical processes occurring at the 
gap between the tip and the substrate surface is important to be understood.

A combination of a STM with an electron microscope have been devised in 
some research groups to see the tip and the substrate surface {\it 
"in-situ"} simultaneously.\cite{gerber,spence,iwatsuki,wada} Those pioneer 
works have successfully revealed the tip-substrate gap \cite{spence} and 
the tip-surface distance.\cite{wada} More recently a plastic change of the 
tip apex was observed during STM operation.\cite{naitoh1}  Furthermore, STM 
tip was used to makea gold nanowire between the tip and the 
substrate.\cite{ohnishi1}

A significant elastic deformation of the tip and the surface is noticed 
under STM operation \cite{chen} or AFM operation.\cite{yamamoto,sugawara} 
Theoretical study on the jump-to-contact demonstrated that attractive 
interaction between the tip and the substrate causes straining of the tip 
and substrate, which sometimes provoke the atom transfer.\cite{guo,grey} 
Although such interaction at atomic level was investigated from force 
measurement by AFM,\cite{perez,cross} contact size and the tip-surface 
distance, have no ways to be detected.

We devised a STM holder attachable to our ultra-high vacuum (UHV) electron 
microscopy,\cite{naitoh1} to observe the tip-substrate contact by 
reflection electron microscopy (REM).\cite{yagi,osakabe} Although the 
deformation of the tip and the surface was small, REM images were sensitive 
enough for detecting strains of the order of $10^{-4}$.

Here, we report REM observation of the \7x7 surface strained by a tungsten 
tip, as the tip approaches to the surface. The strain of the order of 
$10^{-4}$ was first observed to have extended over a circular area of about 
100\,nm.

\section{Experimental}

\subsection{Design of REM holder}

The experiments were performed in the UHV electron microscope (JEM-2000FXV) 
\cite{FXV} whose pressure is better than $5\times10^{-7}$\,Pa.  The REM-STM 
specimen holder (Fig.\ref{holder}) was devised to fit into the narrow gap 
(3.5\,mm) of the objective pole piece. The Si(111) crystal 
(0.02\,$\Omega$cm, n-type) was flash cleaned at 1200\,$^{\circ}$C by 
passing the DC current directly through the Si crystal. Tungsten STM tip 
was sharpened by chemical etching, and preheated for cleaning in an UHV 
chamber before REM-STM experiment. The STM tip was approached to the 
substrate by a mechanical drive (2\,mm) and a stack piezo (6\,\mic{}m), and 
STM image is obtainable by a tube piezo scanner (1\,\mic{}m). The tip 
motion was observed directly by transmission electron microscopy (TEM).

\subsection{STM tip}

Apices of tungsten tips were observed by high-resolution TEM and electron 
diffraction pattern. They always had the (110) plane vertical to the tip 
axis direction and had a curvature of 2\,-\,6\,nm.\cite{naitoh2} Because of 
the preheating before putting the tip into the UHV electron microscope, we 
have not seen heavy contaminations covering over the tip apices. After 
several STM scans on the \7x7 surface, the apices were found often to have 
been scraped. The scanning was usually done at sample bias of $V_{s}< 
2$\,V, and the tunneling current of 0.3\,nA\,$<I_{t}<$\,2.0\,nA. The 
scraped tips were terminated with the (110) plane, which was as wide as 
10\,-\,300\,nm. Such apices of STM tips were used in the following 
experiments. We are allowed observing STM images of the \7x7 surface at 
atomic resolution, when the truncated tip had an adatom cluster on its 
top.\cite{naitoh2}

\subsection{REM imaging of the tip apex}

Ray diagram of the REM is illustrated in Fig.\ref{REMgeo}. The Si(111) 
substrate is placed to the REM-STM holder, whose surface is inclined by an 
angle $\theta_{\rm 0}$ from the objective lens axis. Then the incident 
electron beam is reflected specularly to the surface, propagating along the 
objective lens axis to give the REM image on a fluorescent screen. The 
image is projection of the sample surface, being foreshortened by a factor 
of $\sin\theta_{\rm 0}$ in the direction of the propagating 
beam.\cite{yagi} For the specular beam of the 444 Bragg reflection of the 
Si, the image is foreshortened by a factor of 1/73 
($\theta_{444}=1.36\times{}10^{-2}$\,rad) at the accelerating voltage of 
200\,kV.

When the tip is approached to the surface, the real and its mirror image of 
the tip apex appear on the REM image, as shown in Fig.\ref{REMimage}(a). 
The distance between these apices is $(1+\cos2\theta_{\rm 0})d_{REM}$, 
where $d_{REM}$ is the gap between the tip and the reflection plane of the 
Si(111) surface. Although no RHEED calculation had predicted the position 
of the reflection plane, which should locate between the adatom and the 
stacking-fault layer of the \7x7 surface.\cite{takayanagi}

\subsection{Strain contrast in REM image}

Provided that the surface has no strain, the specular beam changes its 
intensity, $I_0$, for the incident beam angle, $\theta_0$, as illustrated 
in Fig.\ref{straincont}(a).
The rocking curve of the specular beam, a relation of intensity  and 
incident angle , has sharp peaks at the Bragg reflection conditions 
($\theta_0{}=\theta_B$).
The Bragg width, $\Delta\theta_B$, which was the maximum half width of the 
specular reflected beam intensity peak, of the 444 Bragg reflection of the 
Si(111) crystal was calculated to be $7.2\times{}10^{-4}$\,rad on the 
dynamical Bethe theory \cite{yagi2} (100\,kV accelerating voltage was 
assumed in the calculation).
The Bragg width at 200\,kV accelerating voltage is of the same order as 
10$^{-4}$\,rad. When the incident beam angle changes by an amount of the 
Bragg width from the angle of Bragg condition, the REM image changes from 
bright to dark.
Based on this criterion, we understand strain contrast in the REM 
image.\cite{yagi}
When a compressive force is exerted on the surface, surface lattice strains 
as shown in Fig.\ref{straincont}(b).
Provided that the incident beam satisfies the Bragg condition for the 
un-strained surface, the incident beam does not satisfy the Bragg condition 
for the strained area.  Thus, the strained area gives dark contrast, except 
the central area (see Fig.
\ref{straincont}(b)). When the incident angle is smaller (larger) than the 
Bragg angle by the Bragg width for the un-strained area, the dark contrast 
appears only on one side of the strained area.

\section{Results and Discussion}

\subsection{Tip-approach with bias voltage}

The real and mirror images of an STM tip appear in the REM image, as 
reproduced in Fig.\ref{REMimage}(a).
The appearance of the lattice fringes of the $7\!\times\!7$ surface along 
the vertical direction, the [1$\bar{1}$0] direction, proves cleanliness of 
the surface.
The tip is kept at a constant height from the surface, while the STM tip is 
biased by 1.0\,V and tunneling current of 0.8\,nA.
The real and the mirror image gave the tip-surface distance of $d_{\rm 
REM}=1.6\pm{}0.5$\,nm. By further reduction of the voltage 
($-0.3$\,V\,$<V_{s}<$\,0.5\,V, and $I_{t}=0.8$\,nA), the tip approached so 
close to the surface that the separation between the real and mirror tip 
images could not be resolved.
In these bias voltages, a dark horizontal line appeared between the two tip 
images, as shown in Fig.\ref{REMimage}(b).
The dark line extended over an area of 120\,nm. This dark line image is due 
to compressive strain of the \7x7 surface induced by the tip. As explained 
before, contrast analysis confirmed the compressive strain.

Figure \ref{REMtip} shows REM image of a compressive strain, where 
$V_{s}=+0.5$\,V and $I_{t}=0.8$\,nA.
The grazing angle of the incident electron beam increases from (a) to (c), 
passes the Bragg condition in (b).
The strain contrast underneath the tip is dark-bright in (a), dark-dark in 
(b) and bright-dark in (c).
This change indicates the compressive strain.
The strain of the order of $10^{-4}$ extends over 120\,nm, as seen from the 
length of the dark line in Fig.\ref{REMtip}.

The tip-surface distance, $d_{\rm REM}$, was measured as a function of the 
bias voltage ($I_{t}$ is kept constant), and plotted in Fig.\ref{REMtip}. 
The observed $d_{\rm REM}$ vs. bias relationship in Fig.\ref{REMtip} do not 
accord with the previous one that was deduced from the conductance 
oscillation due to the tunneling barrier resonance.\cite{feenstra} $d_{\rm 
REM}$ decreases steeply to zero as the positive bias decreases to $+0.5$\,V 
, or as the negative bias increases to $-0.3$\,V. The bias voltages that 
$d_{\rm REM}$ goes to zero are close to the valence and the conduction band 
edge. When the bias is close to the band edge as in the case of 
Fig.\ref{dvplot}, the tip is almost touching to the sample surface. The tip 
had no mechanical contact, because of repulsive interaction. No trace ofmechanical contact was seen on the \7x7 surface after the retraction of the 
tip from the surface, indeed.

\subsection{Strain of the Si(111) induced by the tip without bias voltage}

The \7x7 surface was also found to be strained by the tip, when no bias 
voltage was applied. The tip was approached to and retracted from the 
surface by the tube piezo scanner. REM images for the tip motion was 
recorded on a videotape, and analyzed in detail. Figure \ref{app-wdraw} is 
a series of REM images, each of which show the real (upper side) and mirror 
(lower side) image of the tip apex. The $7\!\times\!7$ lattice fringes of 
the Si(111) surface were appearing always. As the tip approaches from (a) 
to (d), a dark horizontal line comes out in (b), disappears in (c), and 
reappears in (d). On the way back from (d) to (g), the line contrast 
changes reversibly.
The tip-substrate gap distance was measured in reference to the $d_{\rm 
REM}$ in Fig.\ref{app-wdraw}(a). The $d_{\rm REM}$ in 
Fig.\ref{app-wdraw}(a) was measured directly from the REM to be 1.25\,nm. 
Further approach of the tip did not allow accurate measurement of $d_{\rm 
REM}$ value, so that the gap distance was estimated by $d=d_{\rm 
REM}-\Delta{}d_{\rm piezo}$, where $\Delta{}d_{\rm piezo}$ is the 
elongation of the tube piezo scanner. The gap distance, then, is 0.9\,nm, 
0.4\,nm, 0.15\,nm, 0.45\,nm, and 0.85\,nm for (b)\,-\,(f), respectively. In 
(g), the gap distance became $d_{\rm REM}=1.7$\,nm, which was measured 
directly from the REM image. The length of the dark lines in REM images in 
Fig.\ref{app-wdraw} (and other series of tip approach) were measured as a 
function of the gap distance, $d$, and plotted in Fig.\ref{d-leng}. The 
strain for $0.15\,{\rm nm}<d<0.4\,{\rm nm}$ (Fig.\ref{app-wdraw}(d)) is 
compressive, while that $d>0.45{\rm nm}$ is tensile (Fig.\ref{app-wdraw}(b) 
and (f)). Neither attractive nor repulsive force works at the gap distance 
of $0.45\,{\rm nm}\pm{}0.03\,{\rm nm}$ (Fig.\ref{app-wdraw}(c) and (e)). 
Looking the length of the dark lines (the area having strained more than 
$10^{-4}$\,rad) in Fig.\ref{d-leng}, the range of the strain field is found 
to be extremely wide. The maximum strain field for the attractive 
interaction extends over the area of 50\,nm at the gap distance of 0.8\,nm. 
For repulsive interaction regime, the range extended even more than 100\,nm 
for $d<0.2$\,nm. We calculated the range of the compressive strain field, 
following the classical elastic approach \cite{timo} by assuming various 
radiuses of a flat topped tip and a compressive force . However, no 
reasonable radius or force could explain the magnitude of the strain field 
seen in the REM image.\cite{naitohdr}

In the experiments, we did not observe for our tip to jump-to-contact with 
the Si surface, since our tip is rigid enough. On the other hand, we 
observed jump-to-detach motion of our tip while withdrawal of the tip. On 
the withdrawal, the tensile strain contrast reaches to its maximum in 
Fig.\ref{app-wdraw}(f) at $d=0.85$\,nm. Its contrast is kept constant by 
further withdrawal (0.05\,nm) of the tip, but it disappears suddenly 
(within one frame of the VTR recording: time after 30\,ms). In this 
jump-to-detach motion, the attractive force changed from its maximum to 
zero, and the gap distance had changed from $d=0.9$\,nm to $d_{\rm 
REM}=1.7$\,nm (Fig.\ref{app-wdraw}(g)).
The reason is not clear. The tip apex had no mechanical contact with the 
substrate during the approach and withdrawal process, if the tip-surface 
distance was larger than 0.15\,nm.
When we push the tip to a distance closer than 0.1\,nm, we began to see 
scratch mark on the Si surface after the tip withdrawal.
The tip also strained greatly.
 From these observations, mechanical contact between the tip and the surface 
begins at the gap distances larger than 0.15\,nm.

The present REM-STM observation of the \7x7 surface by a tungsten tip, 
thus, has revealed the strain range of 50\,nm at gap distance of 0.8\,nm. 
Such strain field might cause potential gradient of the surface to excite 
migration of adsorbed atoms or of the surface atoms.\cite{whitman,nakayama} 
The absolute value of the gap distances is supposed to be overestimated, 
since the gap distances (Fig.\ref{REMtip}) in STM condition are larger than 
the previous report.\cite{feenstra} Any way, the strain field of the 
substrate was detected first in this experiment. The strain changed tensile 
to compressive in relation to the attractive and repulsive force from the 
tip, respectively.

Atomic process such as jump-to-contact has not seen in the REM-STM 
experiment. This might be poor resolution of the REM image, and should be 
done by TEM-STM in future. TEM-STM, however, is only sensitive to the 
strain of the order of $10^{-3}$. Detection of the surface strain becomes 
possible by REM imaging of the surface.

\section{Conclusion}

By a combination of STM with UHV electron microscope, the \7x7 surface was 
observed simultaneously in REM and STM. The tip apex could be imaged in REM 
images of the specularly reflected electron beam. From the real and the 
mirror images of the tip, we knew the tip-surface distance directly. The 
tip was approached to and retracted from the sample surface with or without 
bias voltage being applied between the tip and the sample surface. The 
tensile strain was induced on the Si surface when the gap distance is 
$0.45\,{\rm nm}<d<0.85\,{\rm nm}$. The surface strain turns into 
compressive for $\,0.15\,{\rm nm}<d<0.45\,{\rm nm}$. Mechanical contact of 
the tip to the sample surface occur for $d<0.10$\,nm. The range of the 
strain field, the area strained more than $10^{-4}$, is 50\,nm in the 
attractive force regime, while it exceeds 100\,nm in the repulsive 
interaction regime.


%
%
\begin{figure}
\caption{Design features of the STM holder for our electron microscope.
}
\label{holder}
\end{figure}

\begin{figure}
\caption{Schematic illustration of REM imaging. An electron beam with an 
incident angle, $\theta_0$, is specularly reflected by a substrate surface.
The REM image is foreshortened by the factor of $\sin{}\theta_0$.
It is just like the projected image from the virtual incidence.
When the tip is approached to the surface, a true and the mirror images of 
the tip are seen on the REM image of the surface.
The distance between the two tip apices images is given by 
$(1+\cos{}2\theta{}_0)d_{REM}$.
}
\label{REMgeo}
\end{figure}

\begin{figure}
\caption{(a) Schematic illustration for the beam reflection on the surface 
and specular reflection intensity, $I$, as a function of incidence angle, 
$\theta{}_0$. The intensity at the Bragg condition, $\theta_0=\theta_B$, 
has sharp peak with the Bragg width, $\Delta\theta_B$.
(b) Schematic illustration for the changes of incident angle on the 
compressive strained surface and the specular reflection intensitiy 
distribution. Two dark lines contrast at $\theta_0=\theta_B$ change to 
bright-dark (dark-bright) contrast under condition that incident angle, 
$\theta_0$, is larger (smaller) than the Bragg angle, $\theta_B$.
}
\label{straincont}
\end{figure}

\begin{figure}
\caption{(a) shows the REM image of a tungsten tip and a Si(111) surface 
when the tip is approached to the surface in a constant current mode of STM 
($V_s=+1.0$\,V, $I_t=0.8$\,nA and).
The gap distance between the tip and the reflection plane in the substrate 
surface, $d_{REM}$, is estimated to be 1.6\,nm.
(b) is the strained surface image of the Si indicated by a horizontal dark 
line contrast between the two tip images as reducing the sample bias to 
$-0.3\,{\rm V}<V_s<+0.5$\,V.
The dark contrast length implies the strain of the order of 10$^{-4}$ 
exteded over the 120\,nm diameter area.
}
\label{REMimage}
\end{figure}

\begin{figure}
\caption{REM images of the strained surface below the tip at $V_s=+0.5$\,V, 
$I_t=0.8$\,nA taken (a) under out of Bragg condition of 
$\theta_0<\theta_0$, (b) for Si (444) Bragg condition of 
$\theta_0=\theta_B$ and (c) under out of Bragg condition of 
$\theta_0>\theta_B$. Changes of the line contrast between the true and 
mirror tip images are noted.
}
\label{REMtip}
\end{figure}

\begin{figure}
\caption{The $V_s$\,-\,$d$ plot obtained by REM-STM observation. The 
circles and open triangles refer to different tunneling currents, 
$I_t=0.35\,{\rm nA}$ and 0.8\,nA, respectively.
}
\label{dvplot}
\end{figure}

\begin{figure}
\caption{REM images of the straining on a Si surface induced by a tip 
without applying sample bias voltage, which were taken at
(a) $d_{REM}=1.2$\,nm, (b) $d=0.8$\,nm, (c) $d=0.4$\,nm and (d) 
$d=0.1$\,nm in a tip approaching process and at (e) $d=0.45$\,nm, (e) 
$d=0.85$\,nm and (f) $d_{REM}=1.7$\,nm in a tip withdrawing process. }
\label{app-wdraw}
\end{figure}

\begin{figure}
\caption{The contrast length of the strained surface in 
fig.\ref{app-wdraw}(b)\,-\,(f) plotted for the tip-surface distance, $d$.
The length of the compressive (attractive) straining is shown by positive 
(negative) value. }
\label{d-leng}
\end{figure}

%
%

\end{document}